\documentclass[conference]{IEEEtran}
\ifCLASSINFOpdf
\else
\fi
\usepackage{array}
\usepackage{mdwmath}
\usepackage{mdwtab}
\usepackage{eqparbox}
\usepackage{url}

\usepackage{tablefootnote}

\usepackage{amssymb}
\usepackage{graphicx}

\usepackage{xcolor}
\usepackage{multirow}
\usepackage{threeparttable}
\usepackage{hyperref}
\usepackage[all]{hypcap}
\usepackage{amsmath}
\usepackage{array}

\newcolumntype{C}[1]{>{\centering\let\newline\\\arraybackslash\hspace{0pt}}m{#1}}
\usepackage{enumerate}
\usepackage{scrextend}
\usepackage[noadjust]{cite}

\begin{document}
\title{Investigating the Effect of Attributes on User Trust in Social Media}

\author{\IEEEauthorblockN{Jamal Al Qundus}
\IEEEauthorblockA{Computer
Science, Freie Universit\"at Berlin,\\
Email: jamal.alqundus@fu-berlin.de}
\and
\IEEEauthorblockN{Adrian Paschke}
\IEEEauthorblockA{Data Analytics Center, Fraunhofer FOKUS, Berlin,\\
Email: adrian.paschke@fokus-fraunhofer.de}

}
\maketitle

\begin{abstract}
One main challenge in social media is to identify trustworthy information. If we cannot recognize information as trustworthy, that information may become useless or be lost. Opposite, we could consume wrong or fake information - with major consequences. How does a user handle the information provided before consuming it? Are the comments on a post, the author or votes essential for taking such a decision? Are these attributes considered together and which attribute is more important? To answer these questions, we developed a trust model to support knowledge sharing of user content in social media. This trust model is based on the dimensions of stability, quality, and credibility. Each dimension contains metrics (user role, user IQ, votes, etc.) that are important to the user based on data analysis. We present in this paper, an evaluation of the proposed trust model using conjoint analysis (CA) as an evaluation method. The results obtained from 348 responses, validate the trust model. A trust degree translator interprets the content as very trusted, trusted, untrusted, and very untrusted based on the calculated value of trust. Furthermore, the results show a different importance for each dimension: stability 24\%, credibility 35\% and quality 41\%.
\end{abstract}

\IEEEpeerreviewmaketitle


\IEEEpeerreviewmaketitle

\section{Introduction}\label{sec:introduction}
Users consume information when they have trust in it. One main challenge in social media is how to identify trustworthy information. For instance, relevant information such as storm warning or medical instruction could not be considered by users, if it is not recognized as trustworthy. 
Usually, users look at properties (e.g. author, reviews, etc.) to take decision to trust the information. However, many questions arise when it comes to which properties are relevant and how important they are for the user to consume this information. 

Social media (SM) have many users, which makes it well suitable for examining user activities on the information provided. Therefore, we considered the SM Genius (\url{www.genius.com}) as a case study to measure user's willingness to trust the information provided on this platform. 
Interactively, users on Genius create annotations that serve as placeholders for the interpretations of texts, especially lyrics and literature. Annotations provide editing functions such as voting, sharing, adding comments, etc. Participation in this platform is described by certain activities. These activities are linked to certain user authorizations (e.g. roles: whitehat, artist, editor, etc.). 
Based on these authorizations, a user can perform certain activities and earn Intelligence Quotient (IQ)\footnote{Intelligence Quotient is a counter of points awarded for activities on Genius.\label{fn:iq}}, which indicates the experience required for authorization and acceptance of content \cite{qundus2016}.

The focus of this article is to determine the properties that are important for user trust in the user-generated content environment (i.e. any content generated by the user on Genius that is an annotation, a comment, or a modification).

Existing research \cite{li2014effects, miltgen2015exploring, liu2008predicting, adler2011wikipedia} tackles this problem of trust by verifying the history of the generated content, reputation and algorithms for detecting vandalism. In the contract, we estimate user preferences on content properties that are relevant in making decision to consume (trust) the information. This gives us an idea of how we can present information in social media using a template that helps identify trusted information. 

To obtain such a template, the user's willingness to trust should be measured by (1) simulating a number of templates that include different properties and (2) estimating the user's choices. Measuring user's willingness leads to the construction of a trust model that quantifies the value of trust and at the same time embodies in its structure the construction of the required template. This can be evaluated with the use of the conjoint analysis (CA) \cite{LuceTukey1964} method, which simulates the decision-making process of consumers when choosing products in real life. 

To build our model, we analyzed first, the data collected from Genius based on user activities (e.g. annotation, voting, comments). Then, we select the metrics used in Genius that correspond to user activities (for instance, "annotation" corresponds to "annotation IQ\footref{fn:iq}", 
, "vote" to "edit IQ\footref{fn:iq}"). Finally, we looked for a correlation between the metrics defined in the dimensions of the existing trust models in the literature and those defined in Genius. With this literature review, we classified the metrics into three dimensions namely: stability, credibility and quality. These dimensions are then integrated into our trust model. This trust mode can be used as well for other social media having the same content properties.

Our main emphasis in this paper is to provide a reliable assessment of the selected dimensions and their acceptance by web users in general. This can be conducted by estimating the user choices using a Discrete Choice Conjoint analysis (DCC), which is a form of CA evaluation method.

The paper is structured as follows: Section
\ref{sec:relatedwork} provides a brief overview of relevant works. Section \ref{sec:trustmodelconstruction}
describes the dimensions of the trust model and an illustrative example to compute trust. Section
\ref{sec:provision} presents the preparation of the survey. Section
\ref{sec:resultsanddiscussion} reports and discusses the findings. Section
\ref{sec:conclusionandfuturework} contains the summary and concludes with proposals for further investigation.
\section{Related Work}\label{sec:relatedwork}
Dondio et al. propose a Wikipedia Trust Calculator (WTC) consisting of a data
retrieval module that contains the required data of an article. A
factor-calculator module calculates the confidence factors. A trust evaluator
module transfers the numerical confidence value in a natural language
declaration using constraints provided by a logic conditions module
\cite{dondio2006}. This approach refers exclusively to Wikipedia and cannot be
transferred into other domains such as social media. In addition, the aim here
is to detect vandalism and not trust in our definition. The trust model of
Abdul-Rahman and Hailes is based on sociological characteristics. These are
trusted beliefs between agents based on experience (of trust) and reputation
(came from a recommended agent) combined to build a trusted opinion to make a
decision about interacting with the information provided \cite{abdulrahman2000}.
This approach combines reputation and agents together to build trust. These
metrics are not available in our domain.

These two works are closely linked to our work. Our approach is comparable to
that of Dondio et al., and in particular their work has inspired the calculation
of the stability dimension. The authors build the stability
based on the change (defined as edit) in the text length of an article. While
the stability dimension in our work is built on any type of editing (vote,
suggestion, creation, etc.) to an annotation. In addition, we have adapted the trust degree translator from Abdul-Rahman and Hailes \cite{abdulrahman2000}. Based on data base analysis, we manually defined its constraints, that are used for interpretation of the numeric trust value into a human-readable language.

Cho et al. investigated trust in different contexts and discussed trust from
various aspects. The authors employed a survey to investigate social trust from
interactions/networks, and captured of quality of service (QoS) and quality of
information (QoI) depending on a relation between two entities (trustor and
trustee) \cite{cho2015}. However, this is an examination of trust and
reputation systems in online services \cite{jøsang2007}. These works and others
must be placed in a restricted domain to find a relationship between the
communicating entities. However, this is not always possible in an unlimited
domain like social media. Different from the prior works, this survey brings
together the aspects of trust from a specific, but open domain and let entities
evaluate them from outside of this domain. Additionally, we suggest an advanced
equation for measuring trust.
  
\section{Trust Model Construction}\label{sec:trustmodelconstruction}
We conducted investigation on the social media Genius as a case study to build our trust model, as follows: First, we analyzed the data collected from Genius based on user activities (e.g. annotation, voting, comments). Then, we select the metrics used in Genius that correspond to user activities (for instance, "annotation" corresponds to "annotation IQ\footref{fn:iq}", 
"vote" to "edit IQ\footref{fn:iq}"). Furthermore, other metrics are considered such as the metric "author role" (e.g. editor, or whitehat) which is not considered as an activity but as important metric in Genius (for more detail see Genius technical report \cite{qundusGTR2018}). Finally, we looked for a correlation between the metrics defined in the dimensions of the existing trust models in the literature \cite{dondio2006, metzger2003credibility, pranata2016most, warncke2013tell, fogg2001makes} and those defined in Genius.
For instance, the dimension credibility comprised the metrics set: user role, user IQ, attribution and annotation IQ.
With this literature review, we classified the metrics into three dimensions namely: stability, credibility and quality. These dimensions are then integrated into our trust model.

The trust model classifies annotations into four classes \cite{abdulrahman2000} called trust degrees illustrated in Table \ref{tab:trustdegreetranslatro}. These classes were obtained by the database analysis based on the Empirical Cumulative Distribution Function (ECDF) \cite{castro2015empirical}. Next, we present the formulas used to compute trust. 
\begin{table}[!ht]
\centering
\caption{Trust Degree Translator}
\label{tab:trustdegreetranslatro}
\begin{tabular}{| c | c | c | c | c |}

\hline
  Trust Degree & Percentage & Edits number & Edits IQ & User IQ \\
\hline
 vt & 25\% & $>$5 & $>$35 & $>$1000\\
 t & 31.25\% & 2 to 5 & 5 to 35 & 0 to 1000\\
 u& 6.25\% & 0 to 2 & 0 to 5 & -100 to 0\\ 
 vu& 37.5\% & $<$0 & $<$0 & $<$-100\\
\hline
\end{tabular}
\begin{tablenotes}
      \small
      \item {Tab.\ref{tab:trustdegreetranslatro} illustrates the trust degree translator for the
      interpretation of the individual statement trust classes. 		vt = very trusted,
		t = trusted, 
		u = untrusted, 
		vu = very untrusted. 
		Percentage results from the ECDF applied data analysis observed on Genius, and illustrates the distribution of statements' (annotations) trust classes in the data set.
}
\end{tablenotes}
\end{table}
\subsubsection*{Trust}\label{subsubsec:trustcomputation} is
calculated based on the dimensions stability, credibility and quality. The trust value obtained is then interpreted by the trust degree translator to identify the annotation class: very trusted, trusted, untrusted and very untrusted.
\begin{equation}
\begin{array}{l}
trust = \alpha \times stability + \beta \times credibility + \gamma \times quality
\end{array}
\label{equ:trust}
\end{equation}
$\alpha$, $\beta$ and $\gamma$ are the importance factors of each dimension.
\begin{itemize}
\item Stability ($S$) is represented by the annotation edits' distance
(see Equation \ref{dim:stability}), which represents several
content modifications in a time interval. For example, the interval could be between the initial time stamp of the annotation and the current time.
$E(t)$ (see Equation \ref{dim:editsfunction}) specifies the
number of edits at the time stamp t.

\begin{equation}
\begin{aligned}
\{ E(t): t \rightarrow \Phi \thinspace| E:edits function, t:time stamp, \Phi \in \mathbb{Z}\}
\end{aligned}
\label{dim:editsfunction}
\end{equation}
Here, $\mathbb{Z}$ is a set of all integers.
\begin{equation}
\begin{aligned}
\{ S = \sum_{t=t\textsubscript{0}}^{t=p} E(t) &\thinspace| t\& p: time stamp, E: edits function\}
\end{aligned}
\label{dim:stability}
\end{equation}
\item Credibility ($C$) refers to correctness, authorship and
the depth in meaning of information. We consider the type of user
activity on an annotation as editsType. There are complex activities (e.g.
annotation creation) that require agility from the user during execution. We should note that  the ranks of these complex activities in Genius are higher than so-called simple activities (e.g. annotation voting). In addition, we applied a User Credibility Correction
Factor (UCCF), which is calculated based on the users role\footnote{Genius members
have roles that differ in the permissions assigned to them. The IQ numbers earned for an activity also depend on the role type.\label{def:userrole}}, user IQ\footnote{The overall earned IQ count of a user.\label{def:useriq}} and
attribution\footnote{Attribution is the percentage of edits made by a user.\label{def:attribution}} for modifying the credibility status.
\begin{equation}
\begin{array}{l}
C = \dfrac{UCCF + editsTypes}{2}
\end{array}
\label{dim:credibility}
\end{equation}
Where UCCF= foreach author a:  a.attribution $\times$ a.rolePower. 
And a.rolePower = a.role $\times$ a.roleFactor $\times$ a.IQ.
\item Quality ($Q$) is calculated exactly like C, except for the restriction to
n-top active users \footnote{$n$ is a number that the observer can freely select.}, who are ordered based on their attribution.
\begin{equation}
\begin{array}{l}
Q = \dfrac{UCCF' + editsTypes'}{2}
\end{array}
\label{dim:quality}
\end{equation}
\end{itemize}

Before presenting an example that illustrates how the trust is calculated, we introduce the terms utility and importance. 
The utility or part-worth is a measure of how important an attribute level is for a trade-off decision by the user. Whereas the relative importance of an attribute is the delta percentage compared to all utilities. Each time a respondent makes a choice, an accumulator compiles the numbers. These numbers indicate how often a level has
been selected. The algorithm used for calculating the
utilities is logit model\cite{louviere1983}
combined with a Nelder-Mead Simplex algorithm \cite{nelder1965}.

The utility ($U$) of an attribute level ($l$) is calculated by the distance from
its selected value ($S$) to the minimum ($min$) level selected value of the same
attribute ($k$) as shown in the following Equation \ref{equ:utility}:
\begin{equation}
  U_{al} = S_{kl} -  S_{kl_{min}}
\label{equ:utility}
\end{equation}
The importance ($I$) of an attribute ($k$) is calculated by the
difference between the maximum selected number ($S{_kl_{max}}$) and the minimum
selected number ($S{_kl_{min}}$) of this attribute level, divided by the sum of
such differences over all attributes $A=\{a_1, a_2, \ldots, k, \ldots, a_n\}$
(see Equation \ref{equ:importance}).
\begin{equation}
\label{equ:importance}
  I_k = \dfrac{S{_kl_{max}} -  S{_kl_{min}}}  {\sum_{a=1}^{a=n} (S{_al_{max}} -
  S{_al_{min}})} \times 100\%
\end{equation}
\subsection{Illustrative Example}\label{subsec:illustrativeexample}
In this section, we present an example that illustrates how we apply the results of CA in evaluating the dimensions and as a consequence of that, computing trust of an annotation.

\begin{figure}
\centering
\includegraphics[width=0.49\textwidth]{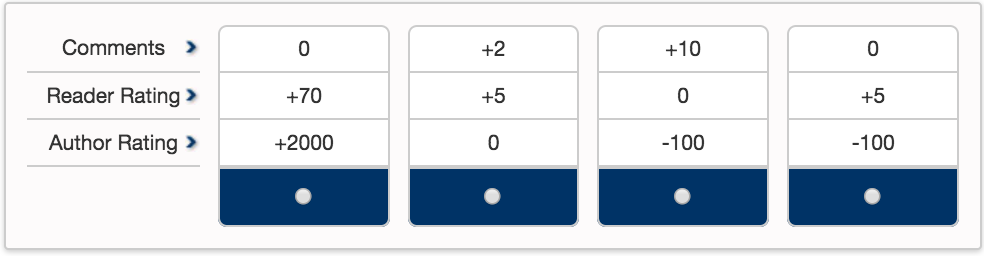}
\caption{\tablefootnote{\url{www.questionpro.com}} presents a task that illustrates one step in the conjoint analysis
profile. This task is displayed to the respondents to make a trade-off on the
provided concepts. A concept consists of attributes (Comments, Reader Rating and
Author Rating) and randomly generated levels values combined to alternatives.}
\label{fig:surveyConcept}
\end{figure}

A discrete choice conjoint analysis (DCC)
provides a quantity\footnote{Number depends on the conjoint analysis design.} of
tasks. A task consists of a set of concepts, each concept represents a certain number
of attribute levels. Users select a concept that they would trust in reality.
Figure \ref{fig:surveyConcept} illustrates a task used in our conjoint design. Each task concept contains the attributes Comments, Reader Rating and Author Rating and their randomly generated levels values. The attributes act in place of the trust dimensions.

Table \ref{tab:attributeImportance} provides an example of the collected numbers of user trade-offs. The columns Level,
Selected and Offered are predefined. We only explain the
calculations for one attribute Comments, since the calculations for the
attributes Reader Rating and Author Rating are analog.
\subsubsection*{\emph{Utility}} Equation \ref{equ:utility}  is applied as
follows: 
$a$ = Comment, 
$l$ = level,
$S_{al_{max}}$ = maximum level selected value,
$S_{al_{min}}$ = minimum level selected value,
$U_{al}$ = 0
\begin{center}
\begin{tabular}{ l| l}
\small
	$l = 0$ & $l = 2$ \\
	$U_{al} = 0 - 0 = 0$ & $U_{al} = 2 - 0 = 2$ \\
    \hline
     	$l = 5$ & $l = 10$\\ $U_{al} = 5 - 0 = 5$ & $U_{al}
	= 10 - 0 = 10$

\end{tabular}
\end{center}

\subsubsection*{relative Importance} (see Equation \ref{equ:importance}) is
applied as follows:\\
  $\sum_{a=1}^{A} (S{_al_{max}} - S{_al_{min}}) = (61 - 2) + (72 - 5) + (80 - 1) = 205$\\
  $ k = Comments$, 
  $S{_kl_{max}} = 61$, 
  $S{_kl_{min}} = 2$.
\begin{center}
$I_k\in\{\alpha,\beta,\gamma\}$\footnote{$\beta$=0.33, $\gamma$=0.39}$ = \dfrac{61 - 2} {205} \times 100\% \approx 29\%$\\
\end{center}

\subsubsection*{trust} ($\omega$) can be now calculated based on the dimensions'
Equations \ref{dim:stability}, \ref{dim:credibility}, \ref{dim:quality} and
\ref{equ:trust} as follows:\\

Let the number of edits of an annotation be 50 from the creation time to the current time. 
From these edits are 10 complex edits (CE) (e.g. content modification) and 40 simple edits (SE) (e.g. voting). The edits IQ equals 30. The users' roles are (editor (25), whitehat (3) and staff (38)), the sum total of users' (authors') IQ equals 160 (10, 30 and 120, respectively) and their attributions (
70\% with 2 CE and 7 SE, 
28\% with 7 CE and 30 SE and 
2\% with 1 CE and 3 SE, respectively). 
The n = 2 for the n-top active user. Based on this input the stability, credibility and quality can be calculated, as follows: 

\textbf{Stability} = $50$ 

\textbf{Credibility} =  $\dfrac{7.285 + 7.5}{2} = 7.392$ 

where UCCF = foreach author a:  a.attribution $\times$ a.rolePower 

And a.rolePower = a.role $\times$ a.roleFactor $\times$ a.IQ 

UCCF = $0.7 \times (25 \times 0.025 \times 10) + 0.28 \times (3\times 0.025 \times 30) + 0.02 \times (38 \times 0.025 \times 120) = 7.285$ 

EditsTypes = edit IQ $ \times \dfrac{CE }{SE}$ = $30 \times \dfrac{10 }{40} = 7.5$ 

\textbf{Quality$_{n=2}$} $=\dfrac{5.005 + 7.297}{2} = 6.151$ 

where UCFF' = $UCFF - 0.02 \times (38 \times 0.025 \times 120) = 5.005$ 

EditsTypes' = editIQ $ \times \dfrac{CE' }{SE'}$ = $30 \times  \dfrac{9}{37} = 7.297$ 

\textbf{Trust} = $0.29 \times 50 +  0.33 \times 7.392 +  0.39 \times 6.151 = 19.338$

\textbf{Trust degree translator} interprets the trust value $\omega \geq  15$ as very trusted, $15 > \omega \geq 13.5$ as trusted, $13.5 > \omega \geq 12$ as untrusted and $\omega < 12$ as very untrusted. 

\begin{table}[!ht]
\centering
\caption{Attributes and Levels Calculation Example}
\label{tab:attributeImportance}
\begin{tabular}{|C{1.65cm} | c | c | c | c | C{1.5cm} |}

\hline
Attribute & Level & Selected & Utility & Offered  & rel.Importance\\
\hline
\multirow{4}{*}{Comment}
 & 0  & 2  & 0  & 5\%  &\multirow{3}{*}{\textbf{29\%}}\\  \cline{2-5}
 & 2  & 33 & 2  & 24\%  &\\  \cline{2-5}
 & 5  & 44 & 5  & 31\%  &\\  \cline{2-5}
 & 10 & 61 & 10 & 40\%  &\\
\hline\hline
\multirow{4}{*}{Reader Rating}
 & 0  & 5  & 0  & 4\%   &\multirow{3}{*}{\textbf{33\%}}\\  \cline{2-5}
 & 10 & 24 & 10 & 17\%  &\\  \cline{2-5}
 & 30 & 39 & 30 & 28\%  &\\  \cline{2-5}
 & 70 & 72 & 70 & 51\%  &\\
\hline\hline
\multirow{4}{*}{Author Rating}
 & -100 & 1	 & 0 	& 5\%     &\multirow{3}{*}{\textbf{39\%}}\\  \cline{2-5}
 & 0 	& 7	 & 100 	& 7\%     &\\  \cline{2-5}
 & 1000 & 52	 & 1100	& 37\%   &\\  \cline{2-5}
 & 2000 & 80	 & 2100	& 51\%  &\\
\hline
\end{tabular}
\begin{tablenotes}
      \small
      \item {Tab.\ref{tab:attributeImportance} gives an example of calculating the attributes relative importance. Attribute = the property of a statement, Level = one possible value an attribute can take, Selected = the number of selection frequency of respondents, Utility = level's important to the respondents choice decision, Offered = display frequency to the respondents, relative Importance = measure of how preferred an attribute is to the respondents choice decision.}
\end{tablenotes}
\end{table}

\section{Methodology}\label{sec:provision}
Our approach applies DCC as follows: Using e-mail, we announced a link to the online survey in Arabic, English and
German. In the DCC we described the attributes (see Table \ref{tab:attributeimportance}): (1) Comments as ``a number
that indicates improvement edits created by other readers'', (2) Reader
Rating as ``a number of other readers' approval'' and (3) Author Rating
as ``a number of voting that the author earned for his activities in the
social network''. We also stated that ``The greater the number, the greater the
satisfaction''. Each number represents the sum of negative and positive
assertions. There are comments that were rated negatively and other comments
that were rated positively by readers. Negatives were marked with minus and
positives with plus numbers. Subsequently, both numbers were summed up. This
applies to all properties''.

Due to the amount of information in a full-profile design
($4_{levels}^{3_{attributes}}$ makes 64 alternatives), a complied questionnaire
would become too extensive. Therefore, we decided to use a fractional factorial-design within the factor $\frac{1}{2}$.  The conducted DCC consists of
32 concepts and the respondents were also given four alternatives to choose from in each task.
\section{Results and Discussion}\label{sec:resultsanddiscussion}
Johnson and Orme recommend a rule-of-thumb for the minimum sample sizes for CBC modeling $(\dfrac{nta}{c} \geq 500)$ \cite{johnson1996}. Where \emph{n}: the respondents number, \emph{t}: the tasks number, \emph{a}: the number of alternatives per task and \emph{c}: the highest level number over all attribute. Accordingly, our questionnaire response has a satisfactory
number of participants \footnote{348 responses with a completion rate of 40.65\%}.
The questionnaire received responses that were distributed over 12 countries, which supports the results to be more significant and we experienced responses from a widely distributed
audience.

The areas of emphasis of the individual statements met all our expectations and
are in line with the proposed theoretical trust model. The selected properties included in the dimensions is important to the users by making a decision to trust the information provided. Table \ref{tab:attributeimportance} presents the best profile (levels: +10, +70 and +2000) and the worst profile ( levels: 0, 0, -100). These two results are interpreted as very trusted and the worst profile as very untrusted respectively by the trust degrees translator. As a reminder, the values of levels in the questionnaire and the numbers applied into the trust degree translator are obtained from the analysis of the database collected from Genius. This table provides information regarding the importance of the attributes and the utilities of the levels. The analysis of the respondents' choice decisions has been conducted by the authors of this paper. This analysis provides a summary from which we can conclude the following:
\begin{enumerate}[I]
\item \emph{None of the attributes was excluded from the choice decision.}\label{fct:i}
All defined dimensions have significance in the proposed trust model. Importance of stability, credibility and
quality are 24\%, 35\% and 41\% respectively)  (see
Table \ref{tab:attributeimportance}). If a dimension had turned out
less significance, this would mean that it has no relevancy for the model and
should not be considered. This result is an indicator that the model is accepted
and confirmed by the evaluators.
\item \emph{None of the attributes has an extremely high value.}\label{fct:ii}
None of the dimensions alone make up the model. That would mean that we can
neglect all other dimensions and focus exclusively on one. In addition, it would
indicate that other components or dimensions are essential for trust in this
context and our model did not consider them.
\item \emph{The importance of the attributes is about equally dispersed.}\label{fct:iii}
This confirms our preliminary consideration when we weighted the dimensions
nearly equal (see Equation \ref{equ:trust}). Nevertheless, it was not possible
to determine a more precise weighting for the trust calculation until this
evaluation. In which, the respondents printed out their importance weight
distribution of each dimension. Applying this weighting into the equation we can
improve the calculation with coefficients that are derived from the percentages
of the individual domain rankings.
\item \emph{There is a distinct subdivision of the utilities of an attribute
into four parts.}\label{fct:iv} The resulted subdivision is in line with the
classes of the trust degrees. The levels of an attribute exhibit clear
differences on how often they have been selected (see Table \ref{tab:attributeimportance}). We can reclassify the distribution of the levels
utilities of individual attribute in the classes. This applies to all attributes.
\item \emph{The distinct subdivisions agree over all attributes respectively.}\label{fct:v}
This corresponds with the prior statement and represents an extension that the
subdivision of the levels utilities on an attribute stage continues to move
across all attributes levels and in the same order (see Table \ref{tab:attributeimportance}). If we number the trust classes and the levels of
each attribute consecutively, we realize that the first level of each attribute
can be assigned to the first trust class (very untrusted) and the second
attributes levels can be assigned to the second class (untrusted) and so forth.
For instance, The utility (-0.90) of the first level (-100) of the attribute
author rating, the utility (-0.82) of the first level (0) of the attribute
reader rating and the utility (-0.67) of the first level (0) of the attribute
comments, present together a concept that was the least rated once by the
respondents. This concept is classified as very untrusted by our trust model.
This is applied for the second, third and fourth levels of each attribute
respectively. 
\end{enumerate}

\begin{table}[!ht]
\centering
\caption{Attributes Importance and Levels Utilities Results}
\label{tab:attributeimportance}
\begin{tabular}{| c | c | c | c | c | c |}

\hline
  \multicolumn{2}{|C{2cm}|}{Author Rating (quality)} & \multicolumn{2}{C{2cm}|}{Reader Rating (credibility)} & \multicolumn{2}{C{2cm}|}{Comments (stability)} \\
\hline
\hline

 \multicolumn{2}{|c|}{40.85\%} & \multicolumn{2}{c|}{34.8\%} & \multicolumn{2}{c|}{24.35\%}\\
\hline
 Level & Utility & Level & Utility & Level & Utility\\
\hline
\hline

 -100 & -0.90& 0  & -0.82 & 0 & -0.67\\
\hline
 	0 & -0.39& +5 & -0.18 & +2& -0.04\\
\hline
 +1000&	+0.44 & +30& +0.29& +5& +0.24\\
\hline
 +2000& +0.86 & +70& +0.71&+10& +0.47\\
\hline
\end{tabular}
\begin{tablenotes}
      \small
      \item {Table \ref{tab:attributeimportance} gives the attributes importance, the levels and their utilities as the measure for the respondents' preferences. This refines the weights of the attributes acting in place of the trust model dimensions.}
\end{tablenotes}
\end{table}

\section{Conclusion and Future Work}\label{sec:conclusionandfuturework}
This work carried out an evaluation of the trust model using a conjoint analysis to determine the
respondents' choice. The information gained from the results confirm our trust
model. In its structure is the required template included that helps identify trusted information according to user estimated preferences. This model consists of three
dimensions: stability, credibility and quality which were adopted from the
literature and the Genius platform. This model is intended to support the
development of successful applications. 

The used logit model provides effective analysis and
convincing results.
Nevertheless, an analysis using the hierarchical bayes would
allow us to take a fresh look at the selections of individuals. With hierarchical bayes we
would be able to trace the history of respondent's decisions
and possibly expose some more details about their behavior. After the
respondents' conclusions about the developed model have been drawn, it is
necessary to investigate the database from which the model was created. A
clustering procedure should be used to identify which text content belong to
which trust category and why. The content-related parameters are to be investigated.

\bibliographystyle{IEEEtran}
\bibliography{trustCABib}
\end{document}